\documentclass[usenatbib,usegraphicx,useAMS]{mn2e}
\usepackage{amssymb,url,times}
\usepackage{epsfig}
\usepackage{natbib}

\newdimen\figurewidth
\def\mr{\mathrm}
\def\gtrsim{\mathrel{\hbox{\rlap{\hbox{\lower3pt\hbox{$\sim$}}}\raise2pt\hbox{$>$}}}}
\def\lesssim{\mathrel{\hbox{\rlap{\hbox{\lower3pt\hbox{$\sim$}}}\raise2pt\hbox{$<$}}}}
\def\be{\begin{equation}}
\def\ee{\end{equation}} 
\def\no{\noindent}

\def\Aconst{\varepsilon}

\title[Growing Supermassive Black Holes seeds]{A Lower Limit on the Halo Mass to form Supermassive Black Holes}
\author[Dotan et al.]{Calanit Dotan\footnote{calanit@phys.huji.ac.il}$^{1}$, Elena M. Rossi$^{1}$$^{,2}$ \& Nir J. Shaviv$^{1}$ \\
$^{1}$Racah Institute of Physics, The Hebrew University, Jerusalem 91904, Israel \\
$^{2}$Leiden Observatory, Leiden University, P.O. Box 9513, 2300 RA , Leiden, The Netherlands }
\begin{document}

\pagerange{\pageref{firstpage}--\pageref{lastpage}} \pubyear{2010}

\maketitle
\begin{abstract}
We consider a scenario where supermassive black holes form through 
direct accumulation of gas at the centre of proto-galaxies. 
 In the first stage, the accumulated gas forms a super-massive star whose core collapses
when the nuclear fuel is exhausted, forming a black hole of $M_{\rm BH} \approx 100 M_{\sun}$.
 As the black hole starts accreting, it inflates the surrounding dense gas into an almost hydrostatic 
self-gravitating envelope, with at least $10-100$ times the mass of the hole.
 We find that these ``quasistars" suffer extremely high rates of mass loss through winds from their envelopes,
 in analogy to very massive stars such as $\eta$-Carinae.
 Only for envelope masses greater than $2.8 \times 10^{5} (M_{\rm BH}/100 M_{\sun})^{9/11}$ 
is the envelope evaporation time-scale longer than the accretion time-scale of the black hole. 
This relation thus constitutes a ``threshold growth line" above which quasistars can grow their internal black holes. 
Accretion rates can be $10$ to $100$ times the Eddington rate. The quasistars born in this ``growth region" 
with $10^7-10^8 M_{\sun}$  can grow black holes with masses between $10^4$ to $10^5 M_{\sun}$, 
before crossing the threshold growth line and dispersing their envelopes in less than $10^4$ yr. 
 This scenario therefore predicts that massive black hole seeds can be found only in dark matter halos with total 
masses larger than about $10^9 M_{\sun}$, which can provide sufficiently high accretion rates to form such massive
quasistars.
\end{abstract}


\section{Introduction}
\label{sec:intro}
The formation of supermassive black holes (SMBHs) requires the inflow
of $ > 10^6 M_{\sun}$ of gas from galactic scales into a region much smaller than a parsec.
The presence of bright ($\sim 10^{47}$ erg s$^{-1}$) quasars at $z >6$ \citep{Fan2001} suggests
that, at least in those cases,  this inflow must have started at an earlier 
epoch and proceeded at sufficiently high rate to have allowed the assemblage of 
black holes (BHs) of $10^9 M_{\sun}$, in less than a Gyr.  In proto-galactic dark
matter halos with mass $> 10^9 M_{\sun}$, gas can be funnelled towards the centre
at a rate of $> M_{\sun}$ yr$^{-1}$.

 If substantial
fragmentation of gas into stars can be avoided \citep[e.g.,][]{Wise2008}, the collection
of gas at such high rates can lead to the formation of a light
($\lesssim 100 M_{\sun}$) black hole (BH) via direct collapse of gas or through 
the intermediate stage a supermassive star, whose core collapses to form a BH \citep{Begelman2010}.
The feedback
from this ``embryo" BH becomes important and it modifies the structure
of the pregalactic discs in its innermost $\approx 100$ AU. The BH
luminosity inflates the highly opaque gas in its vicinity into a
pressure supported envelope, at least $10-100$ times more massive than the BH
itself. While the envelope is being fed by the proto-galactic disc,
the BH accretes from the envelope. 
This structure is dubbed a {\it quasistar}, since it resembles
a (scaled-up) red giant in structure, though it is powered by accretion into a
central BH \citep[][hereafter BRA08]{Begelman2006,Begelman2008}.

BRA08 found
that for each black hole mass $M_{\rm BH}$, there is a minimum envelope mass $M_*$ below which
no hydrostatic solutions for the whole envelope can be obtained. This line corresponds to modest $M_*/M_{\rm BH}$ ratios of a few tens. Above this line,
 the BH accretes at the Eddington rate for the whole mass ($M_{\rm T} = M_{\rm BH}+M_*$). The BH can accrete at a super-Eddington rate
(for the BH mass), because the energy released by the accretion is
transported outwards through convection (and not by radiative
transfer).  After a few Myr, the BH reaches a maximum mass of
$\sim 10^4 M_{\sun}$,  which provides a luminosity that unbinds the
surrounding envelope. This physical picture has been confirmed by numerical simulations (Ball et al. 2011, hereafter BTZE11).
Accretion will then proceed at a more ``modest" 
rate through the proto-galactic disc. Accretion and mergers with other holes allow these ``BH seeds" to evolve over cosmic times into SMBHs.
 
An important characteristic of quasistars is that they are loosely
bound, since they are radiation pressure dominated.
A luminosity that
is close to the Eddington limit therefore has the potential to drive 
strong radiation-driven outflows.  The presence of an outflow, in
turn, would modify the quasistar's internal structure, the BH
accretion rate and the final mass of the BH seed that can be produced.
Quantitatively, however, the extent and consequences of such winds are
currently not known. 

In this paper, we aim to address the above
questions. We calculate approximate solutions for a hydrostatic massive
envelope powered by accretion, allowing for mass loss from its
surface. We find that the presence of a wind allows for the existence of hydrostatic
solutions for the envelope, even if the system is super-Eddington (for the total mass). 
With the wind, there is a competition between accretion and evaporation, and
the BH can be effectively fed and increase its mass only when the
envelope evaporation is sufficiently weak. This happens for very
massive envelopes $> 10^6 M_{\sun}$, whose wind reaches a maximum
mass loss, driven by all the available luminosity.

We find that under the above conditions, a BH seed of $> 10^3$ solar masses can be assembled in less than $10^4$ yr.
In particular, to obtain BH seeds of $10^4-10^5 M_{\sun}$, an envelope of mass $>10^7 M_{\sun}$ is required. 
We will show that, as a consequence, the present scenario predicts the presence of massive BH seeds at high redshift ($z \gtrsim 10$) {\em only} in
halos of least $\approx {\rm a \ few} \ 10^{9} \,M_{\sun}$.

The paper is organized as follows. In \S\ref{sec:analytics}, we describe our model and carry out analytical estimates. 
In \S\ref{sec:numerical}, we describe the numerical solution for the structure of super-Eddington quasistars with winds, and we present our results in \S\ref{sec:results}. We discuss the implications of our findings to the modeling of supermassive BH formation
in \S\ref{sec:discussion}.

\section{The ``quasistar" Structure}
\label{sec:analytics}
\begin{figure*}
\includegraphics[width=0.8\textwidth]{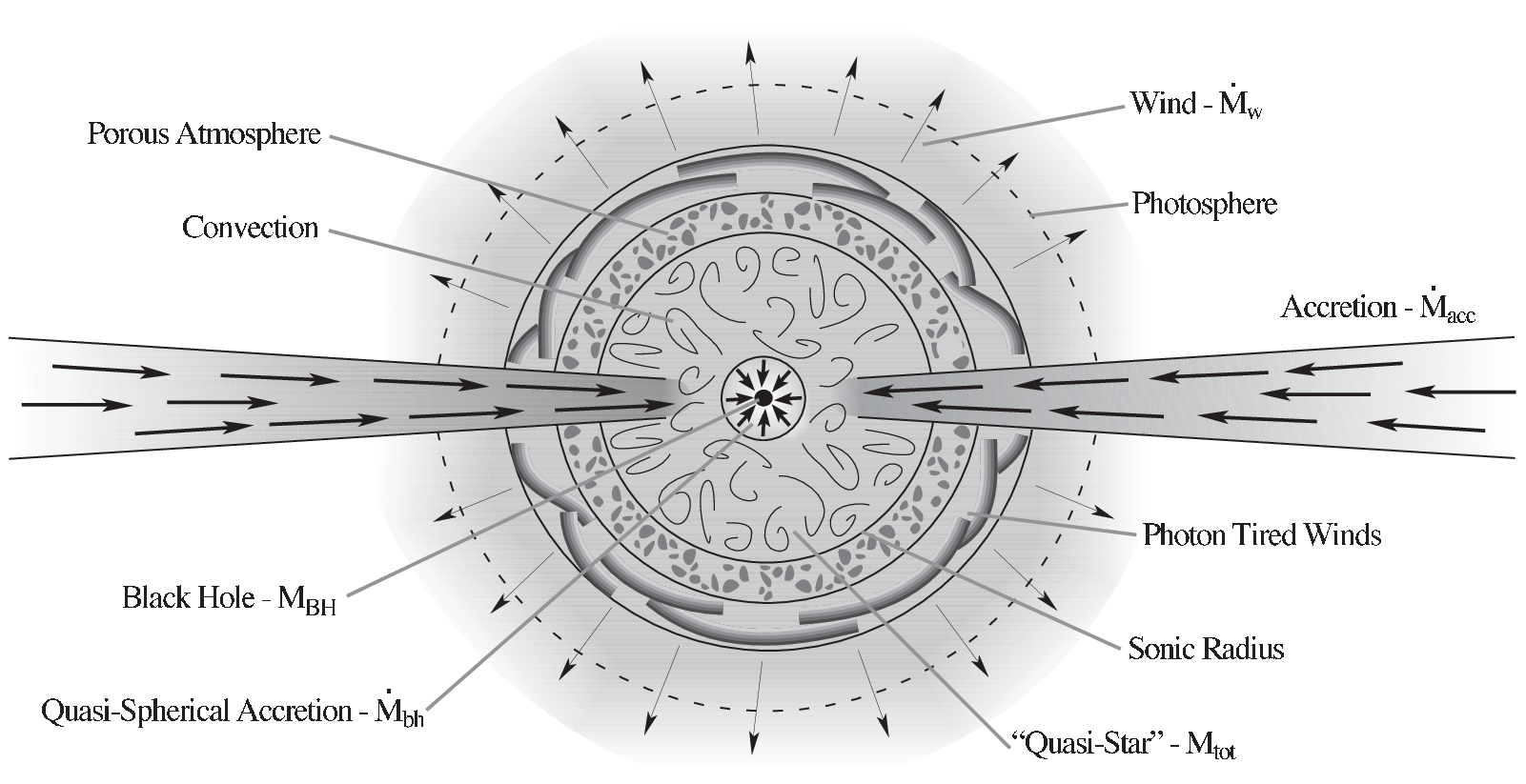}
\caption{A heuristic description of the system we solve. See detailed description in \S\ref{sec:analytics}.}
\label{fig:cartoon}
\end{figure*}

The system we consider is heuristically illustarated in Fig.~\ref{fig:cartoon}.
Crudely, it comprises of five primary regions. From the inside out they are:  
\begin{enumerate}
\item A core, where an embryo BH is fed through a convectively dominated accretion disc.
\item A convective envelope (which contains most of the quasistar mass). 
\item A porous atmosphere, where convection is inefficient.
\item An optically thick wind. 
\item A geometrically thin (at least on these scales) proto-galactic disk. 
\end{enumerate}
The first four regions comprise the quasistar, the structure of which we solve here.

The last region is that of the disk which feeds the quasitar. 
 For reasonable galactic accretion rates,
less than a few tens of  $M_{\sun}$ yr$^{-1}$  the disc is locally sub-Eddington at the radius of the quasistar\footnote{
Using the results of \S2.2 for the radius $r_*$ and mass $M_*$ of a $n=3$ polytropic star, one can show 
that for $\dot{M} < 50$ $M_{\sun}$ yr$^{-1}$, the accretion will release less than $L_{\rm Edd}$, that is,  $\dot{M} \lesssim L_{\rm Edd}/(GM_*/2 r_*)$.}.
Since the disc remains slim, we can assume that mass exchange between the disc and the quasistar happens on a narrow equatorial region.
Moreover, the accreted mass is probably redistributed within the quasistar over a hydrostatic timescale, which is, as we will comment later, much shorter 
than both the accretion and the evaporation timescales.
Thus, we can treat the disc just as a mass source term 
for the quasistar mass budget. Only in one case considered here, for the extreme rate of 300 $M_{\sun}$ yr$^{-1}$,  can the disc be geometrically thick and the interaction between the accretion and the wind
from the quasistar surface may be dynamically important. Therefore,  the results obtained in this case have to be regarded as indicative only.

In the following sections, we describe in more detail the different components and construct an analytical model that
will allow us to delineate the region in the $M_*-M_{\rm BH}$ parameter space, where it is possible to grow massive BHs. 
These BHs can then serve as seeds for the SMBHs observed at the centre of galaxies.
First, however, we will state and justify the primary assumptions we employ to construct our model.

For our purposes, it is not necessary to model the hydrodynamics of the innermost accretion flow,
which by radial extension and mass is a negligible fraction ($\sim10^{-5}$) of the quasisitar.
The theoretical luminosity expected from this ``hypercritical" regime of accretion
is used as a source term in the energy equation for the envelope. Nevertheless, we do consistently consider the accretion rate which in turn depends on the envelope structure, as we explain in the next section.

It has been shown that numerical modeling of quasistars is sensitive to the  choice of the inner envelope radius (BTZE11). 
However, in absence of a physically sound alternative, the most natural choice remains the Bondi radius of the inner accretion disc.
We therefore adopt it,  as it facilitates the analytical calculations and the comparison of our results with previous numerical studies.

 It is possible to show analytically and numerically that a hydrostatic envelope exists
only if it is much more massive than the BH (BRA08, BTZE11).
Therefore, we seek solutions satisfying this condition. 
On the other hand, there is an upper limit on the mass of quasistars, given by stability considerations.
Such radiation dominated structures (with an almost null total binding energy) are in fact subject to the pulsation instability.
\cite{Fowler1966} showed that even a small degree of rotation can stabilize
 stars with mass $<10^{8} M_{\sun}$. Although the envelope is mainly pressure supported, quasistars are indeed expected to be (slowly) rotating since they are fed by pre-galactic accretion discs.
 We therefore assume $M_* \sim 10^{8}$ as an upper limit for the envelope mass.
 
Rotation would also slightly squash the envelope at the poles. Nevertheless, 
it is not trivial to predict the exact rotational profile expected inside the quasistar.
 As we shall see, the quasistar will be almost entirely convective.
 We can therefore expect an azimuthally dependent rotation rate, as exhibited in the sun. 
However, this behavior is still not fully understood. 
As a consequence, we will model the quasistar while neglecting the effects of rotation, 
that is, while assuming spherically symmetry, as in previous works.   

\subsection{The core}

In the innermost region (hereafter ``the core", in analogy to stars), the embryo BH is fed
through an accretion disc which transfers mass from the surrounding hydrostatic and much more massive envelope. 
The extent of the core is the sphere of gravitational influence of the BH, the radius of which is the Bondi radius $r_{\rm b} = GM_{\rm BH}/2 c_{\rm c}^2$, where the local adiabatic sound speed 
 is $c_{\rm c} \equiv  c_{\rm s}(r_{\rm b}) =\sqrt{4P_{\rm c}/3 \rho_{\rm c}}$. Here $\rho_{\rm c}$ and $P_{\rm c}$ are respectively the matter density ($\rho$) and total pressure ($P$) at that radius.

The small angular momentum in the system is such that very close to the last stable orbit, the flow is rotationally supported and viscous processes
allow the release of a luminosity of $L_{\rm BH} = \eta \dot{M}_{\rm BH} c^2$, with a radiation efficiency of $\eta\equiv \eta_{0.1} \times 0.1$. At larger radii, the disc has a ``thick"   geometry 
(\citealt{Narayan1994}; \citealt{Abramowicz1995}; \citealt{Blandford1999}), which is radiatively inefficient due to its high density (\citealt{Begelman1979}; \citealt{Begelman1982}; \citealt{Blondin1986}). Since the radial convection can be shown to be important, the innermost flow is therefore modeled as a 
convection dominated disc (CDAF, \citealt{Stone1999}; \citealt{Igumenshchev1999}; \citealt{Quataert2000}).

At steady state, all the forces in these discs scale with gravity. Thus, $c_{\rm s} \propto r^{-1/2}$ and a ``modified hydrostatic" balance holds.
Moreover, contrary to classical advection dominated discs, CDAFs are characterized by a net (and constant) outward flux of energy, 
which is possible only if $\rho \propto r^{-1/2}$
(\citealt{Stone1999}; \citealt{Quataert2000}). 

These scalings imply that the maximum luminosity that convection can transport within the flow,

\be
L_{\rm c,max} = 4 \pi r^2 \rho c_{\rm s}^3,
\label{eq:lmax_c}
\ee

\no
is constant with radius. This limit for the carrying capacity of convection is set by the requirement that the convective cell motion should remain subsonic. That is, the convective velocity can be at most the local sound speed $c_{\rm s}$, otherwise shocks will form and dissipate the motion over distances much shorter than the pressure scale height. 
Therefore, the energy released at small radii can be transported by convection up to $\simeq r_{\rm b}$, if  $L_{\rm BH} \le L_{\rm c,max}$.
We thus assume 

\be
L_{\rm BH} = \alpha L_{\rm c,max},
\label{eq:Lbh_num}
\ee

\no
with $\alpha \le 1$. The parameter $\alpha$ not only accounts for convective transport inefficiency but also for possible leakage of energy (e.g., by jets close to the hole) which would decrease the supply of energy into the envelope.

Since $L_{\rm c,max}$ is constant in the flow, it can be indeed evaluated anywhere not too close to the BH. However, for reasons that will be clear in the next section, it is convenient to  
express $L_{\rm c,max}$ at $r_{\rm b}$,

\be
L_{\rm BH} = \alpha \sqrt{\frac{9}{4 a}} \pi (G M_{\rm BH})^2 \rho_{\rm c}^{3/2} T_{\rm c}^{-2},
\label{eq:Lbh}
\ee

\no
where we took into consideration that this hot flow is radiation pressure dominated, $P_{\rm c} \simeq a T_{\rm c}^{4}/3$,
where $a$ is the radiation constant. 

\subsection{The hydrostatic region}
\label{sec:analytic_envelope}
In a quasistar, most of the mass and the radial extent are occupied by a highly convective envelope.
The convection efficiently transports the accretion energy flux to large radii ($\approx 100$ AU) where it powers a wind,
after having crossed a thin hydrostatic radiative layer. 
In the following estimates, we will assume that the mass and the radius are constant in this radiative layer and equal to those of the entire convective envelope ($M_*$ and $r_*$ respectively).    

The structure of a strongly convective hydrostatic envelope can be described by an $n=3$ polytrope (while employing the classic Lane-Emden solution).
These massive stars have a small fixed ratio between the gas and radiation pressure of 

\be
\bar{\beta} \approx 7 \times 10^{-3} (M_*/10^6 M_{\sun})^{-1/2} \ll 1.
\label{eq:beta}
\ee
Since the central density and pressure profiles are rather flat with radius, we can use the central 
polytropic values to estimate the density and temperature at the Bondi radius, and thus evaluate $L_{\rm BH}$ (as given by eq. \ref{eq:Lbh}).

Scaling the envelope mass as $M_* = m_* \,M_{\sun}$ and the central temperature as $T_{\rm c} =  T_6 \ 10^6$ K, the envelope radius is given by 

\be
r_* = 5.8 \times 10^{12} m_*^{1/2} T_6^{-1} \;\;{\rm cm},
\label{eq:rs_n3}
\ee

\no
while the central density is

\be
\rho_{\rm c} = 1.3 \times 10^{-4} m_*^{-1/2} T_6^{3} \;\; {\rm g \; cm^{-3}} 
\label{eq:rhoc_n3}
\ee

\no
(e.g., see \citealt{Hoyle1963}).

Furthermore, \cite{Joss1973} have shown that wherever the density is high enough, the local luminosity can never exceed the local Eddington limit $L_{\rm edd,r} = 4 \pi c G M(r)/\kappa(r)$, as calculated with the enclosed mass $M(r)$, and the local opacity $\kappa(r)$. This is because convection will be excited, and it will advect a large enough flux to keep the system sub-Eddington. 
However, there is an upper limit to the luminosity that can be transported by convection at each location (eq.~\ref{eq:lmax_c}).
Outside the Bondi radius, the density and temperature are nearly constant such that $L_{\rm c,max} \propto r^{2}$. Therefore, the convective carrying capacity increases with radius near the centre. At larger radii however,  both $\rho$ and $T$ decrease steeply, forcing a progressively higher fraction of the luminosity to be transported by diffusive radiative transfer. Eventually $L_{\rm BH} =L_{\rm c,max}$. This location, which we call the ``transition radius" ($r_{\rm tr}$), marks the base of the radiative layer.
We search for solutions for which this region is in hydrostatic equilibrium, or in other words, solutions in which a radiative atmosphere is present.

As atmospheres  approach $L_{\rm edd,r}$ a plethora of instabilities can set in, whether 
the atmospheres are only Thomson scattering \citep{ShavivInstabilities}, whether they
have magnetic fields \citep{Bubbles1,Bubbles2,Bubbles3}, or even complex 
opacity laws \citep[][which are probably not relevant for the primordial gas in quasistars]{Opacity1,Opacity2}. Once the instabilities grow to 
become nonlinear, they create low density channels for the photons to escape, and effectively suppress the radiative force exerted on the gas \citep{ShavivPorous}.
Such a reduction in the opacity is the only known way to explain the super-Eddington nature of classical novae and $\eta$-Carinae \citep{ShavivEtaCar,ShavivNovae}.
Although we are far from having a complete understanding of the nonlinear state, we can account for it with a reduced opacity 
$\kappa_{\rm eff}$  \citep[e.g.,][]{Dotan2010}. The functional form of this {\em effective} 
opacity $\kappa_{\rm eff}$ will be further elaborated in \S\ref{sec:radiative layer}. 
The Eddington ratio is defined as $\Gamma \equiv L/L_{\rm edd,r}(\kappa)$, where $L(r)$ is the radiative luminosity. 
In the radiative atmosphere $L(r)= L_{\rm BH}$. 

With a reduced opacity it becomes possible to have a static radiative layer, 
if we further impose the approximate condition that the luminosity at the transition radius equals the {\em effective} Eddington limit, then


\be
 L_{\rm BH} = 1.4 \times 10^{38} \; \tilde{\kappa}_{\rm tr}^{-1}\, m_* \; {\rm erg \; s^{-1}},
 \label{eq:lbh_ledd}
\ee

\no
where $\tilde{\kappa}_{\rm tr}$ is  the  effective opacity at the transition radius ($\kappa(r_{\rm tr})= \kappa_{\rm tr}$), 
normalized to the electron scattering value, $\kappa_{\rm es}=0.35$ cm$^2$ g$^{-1}$. 

In the radiative atmosphere, the luminosity satisfies $L_{\rm BH} \lesssim L_{\rm edd,r}(\kappa_{\rm eff})$,
up to the radius where the gas clumps become optically thin such that they cannot funnel the radiation anymore into the rarified regions. 
As the clumps become transparent, the opacity therefore recovers its microscopic value 
($\kappa_{\rm eff} \rightarrow \kappa$) and $L_{\rm edd,r}(\kappa_{\rm eff}) \rightarrow L_{\rm edd,r}(\kappa)$, 
implying that above the radiative atmosphere, there is a radius 
 beyond which no hydrostatic solution can exist. This causes the quasistar to lose mass through a wind.
In the whole section, we will assume that the wind sonic point is close to $r_*$, i.e. $r_{\rm s} = r_*$ (see section \ref{sec:wind} for a detailed discussion of the wind solutions).

 Following BRA08, we combine eq.~\ref{eq:Lbh} and eq.~\ref{eq:rhoc_n3}, and by equating the two expressions for $L_{\rm BH}$ (eqs.~\ref{eq:Lbh} and \ref{eq:lbh_ledd}),
 we derive an expression for the central temperature:
 \be
 T_{\rm c} \approx 1.7 \times 10^4 \; (\alpha \tilde{\kappa}_{\rm tr}\,)^{-2/5} \, m_{\rm BH}^{-4/5} \, m_*^{7/10} \;\; {\rm K}.
 \label{eq:tc}
 \ee
 \no 
If we insert it into eq.~\ref{eq:rs_n3} and eq.~\ref{eq:rhoc_n3}, we finally get
 
 \be
 r_* \approx 3.5 \times 10^{14} \, \left( \alpha \tilde{\kappa}_{\rm tr}\right)^{2/5}\, m_{\rm BH}^{4/5} \; m_*^{-1/5} \;\; {\rm cm},
 \label{eq:rqs}
 \ee
 \no
 and
 
 \be
 \rho_{\rm c} \approx 6.4 \times 10^{-10} \; \left( \alpha \tilde{\kappa}_{\rm tr}\right)^{-6/5}\; m_{\rm BH}^{-12/5} \, m_*^{8/5} \;\; {\rm \frac{g} {cm^{3}}},
 \ee
 
\no
respectively.

 \subsection{The ``no-solution line''}
 
 We now proceed to calculate the line, in the $M_{\rm BH}-M_*$ parameter space,
 below which no hydrostatic envelope is possible. We term this boundary as the 
 {\em the no-solution line}. As we shall see, it corresponds to  low envelope masses ($M_*/M_{\rm BH} \approx 10$). 
 
Without mass loss, solutions with progressively smaller total envelope masses 
(for a given BH mass) have progressively lower photospheric temperatures (BRA08). We find numerically (section 3) that the presence of a wind does not alter this conclusion. As the envelope mass decreases, the photospheric temperature will eventually reach the recombination temperature, below which the opacity drops precipitously. Any further decrease in the envelope mass reduces drastically the photospheric opacity, causing the photosphere to reside further inside the wind, where densities are higher. This continues until the photosphere reaches the base of the 
 wind, that is, $r_{\rm ph} \simeq r_{\rm s}$. 
When the envelope mass is small enough to allow this to occur, 
the photospheric temperature will be given by $\sigma T^4_{\rm ph} \simeq L_{\rm BH}/(4 \pi r_*^2)$, which can be evaluated to be 
 
 \be
 T_{\rm ph} = 1.1 \times 10^3 \, \tilde{\kappa}_{\rm tr}^{-9/20}\; \alpha^{-1/5}\; m_{\rm BH}^{-2/5}\;\;m_*^{7/20}\; {\rm K},
 \label{eq:tph}
 \ee

 \no
 where $\sigma$ is the Stefan-Boltzmann constant. 

When $r_{\rm ph} \simeq r_{\rm s}$, the presence of an entirely optically-thin  wind does not affect 
the structure of the hydrostatic envelope and we should recover the solutions derived for a purely 
hydrostatic quasistar. Under that assumption,  
BRA08 have shown that---in analogy to the Hayashi track for red giants and protostars---no hydrostatic 
solution can be found below a minimum photospheric temperature $T_{\rm min}$, which is just a factor of 
2 lower than the recombination temperature. 
It was shown that $T_{\rm min}$ varies at most by a factor of two around $4500$ K for BHs in the range 
between $M_{\rm BH}=1$ to $10^4$ $M_{\sun}$ (see their fig. 3). 

Assuming a constant floor temperature 
and solving for $T_{\rm ph} = T_{\rm min}$, we find that  the ``no-solution line" is given by
  
 \be
 m_* = 1.1 \times 10^{4} \, \alpha^{4/7} \, \tilde{\kappa}_{\rm tr}^{9/7}\,\left(\frac{T_{\rm min}}{4500 \, {\rm K}}\right)^{20/7}\, \left(\frac{m_{\rm BH}}{100}\right)^{8/7}.
 \label{eq:nsl}
 \ee
   
We note here that, while BRA08 link the existence of this limit to a steep drop in opacity in the radiative layer,
BTZE11 argue that it is a general feature of polytropic models and not one linked to the adopted opacity law.
Both papers however, agree that this limit exists and that it corresponds to $M_*/M_{\rm BH} \approx 10$, which is what we obtain with
eq. \ref{eq:nsl} (see Fig. 2).

\subsection{The wind and the evaporation strip}

It should be stressed that even if a hydrostatic envelope can exist, it does not guarantee that
 a quasistar can ``live" long enough for the BH to grow inside it. In the presence of strong winds, 
we should also require that the accretion time-scale,

\be
t_{\rm BH} \equiv \frac{M_{\rm BH}}{\dot{M}_{\rm BH}} = \frac{M_{\rm BH}}{L_{\rm BH}/\eta c^2},
\ee
\no
is shorter than the evaporation time-scale,

\be
t_{\rm ev} \equiv \frac{M_*}{\dot{M_{\rm w}}} \simeq \frac{G M_*^2}{\Aconst \,L_{\rm BH} r_* }.
\ee


\no
Here, we normalized the mass loss rate to the maximal possible rate, 
obtained when all the accretion luminosity is used to drive the wind out of the 
gravitational potential well, $\dot{M}_{\rm w,max} = L_{\rm BH}/(v^2_{\rm esc}/2) \approx L_{\rm BH} \,r_*/ G M_*$,

\be
\dot{M}_{\rm w,max} \approx 14.8 \, \alpha^{2/5}\,  \tilde{\kappa}_{\rm tr}^{-3/5} \left(\frac{m_{\rm BH}}{100}\right)^{4/5}\,\left(\frac{m_*}{10^6}\right)^{-1/5}\; \rm{M_{\sun}\, yr^{-1}}.
\label{eq:mdotwind_max}
\ee
where $v_{\rm esc}$ is the escape velocity. 
That is, we have written that

\be
 \dot{M}_{\rm w} \frac{v_{\rm esc}^2}{2} \simeq \Aconst \, L_{\rm BH},
\label{eq:mdotwind}
 \ee
\no
with an efficiency factor $\Aconst \equiv \dot{M}_{\rm w}/\dot{M}_{\rm w,max} \le 1$. 

Using the expression for the quasistar radius (eq.~\ref{eq:rqs}) and $\eta=0.1 \,\eta_{0.1}$, we get

\be
\frac{t_{\rm ev}}{t_{\rm BH} } \approx 4.2 \times 10^{-9}  \; \frac{m_*^{11/5} m_{\rm BH}^{-9/5}}{\Aconst\, (\alpha\, \kappa_{\rm tr})^{2/5} \, \eta_{0.1}}.
\label{eq:ration_t}
\ee
\no
The above ratio, at the no-solution line (eq.~\ref{eq:nsl}), is smaller than unity for $M_{\rm BH} < 10^4 M_{\sun}$,
even if we assume that only $1$ percent of the luminosity has been used to accelerate the wind (i.e., $\Aconst = 10^{-2}$).

Let us now fix $M_{\rm BH}$ and explore what happens as we increase the envelope mass.
First, we note that the strongest dependence is on the envelope mass. Therefore, as we increase
 $M_*$, $t_{\rm ev}/t_{\rm BH}$ increases and it will eventually be greater than unity.  
Also, the radius of the star $r_*$ {\em decreases} and the depth of the potential well from which 
matter should escape increases: $v^2_{\rm esc} \propto M_*^{6/5} / \kappa_{\rm tr}^{2/5}$.

Since the mean density $M_*/r_*^3$ increases as well, it is reasonable to assume that, 
as the envelope mass increases, the radial distance between the sonic radius and the photospheric radius increases.
 This implies that an increasing fraction of the luminosity is used to pull the gas out of the potential well (i.e., $\Aconst \rightarrow 1$). 
At that point,  $\dot{M}_{\rm w} \simeq \dot{M}_{\rm w,max}$.

We anticipate here one of our numerical result (elaborated in \S\ref{sec:results}), that the condition $t_{\rm ev}/t_{\rm BH}  > 1$  
is found above the $\epsilon = 1$ limit for all BH masses we consider ($M_{\rm BH} < 10^{5} M_{\sun}$).
Therefore, the equal evaporation/accretion line is  given by

\be
m_*\approx 2.8 \times 10^5 (\alpha \tilde{\kappa}_{\rm tr})^{2/11} \eta_{0.1}^{2/11} \,\left(\frac{m_{\rm BH}}{100}\right)^{9/11}.
\label{eq:tev=tacc}
\ee

 There is a relatively wide strip in the $M_*-M_{\rm BH}$ parameter space between the no-solution
 line (eq.~\ref{eq:nsl}) and the equal-time-scale line (eq.~\ref{eq:tev=tacc}), that we dub {\em the evaporation strip}. 
 Only above this strip is it  possible ``to grow" black holes inside quasistars. Therefore, we will refer to the 
$t_{\rm ev}/t_{\rm BH} =1$ line as {\em the threshold-growth line}, and to the region above as {\em the growth region}.

\subsection{The quasistar Evolution}
\label{sec:analytic_evolution}
 Let us now assume that the envelope is constantly fed at a rate of a few $M_{\sun}$ yr$^{-1}$ from the pregalactic disc.
From eq.~\ref{eq:mdotwind}, we deduce that, in fact, the mass of the envelope will likely {\em not} increase, since the mass loss will (over)compensate the feeding. 
An upper limit for the BH mass that can be reached is therefore obtained by setting the envelope mass to be constant in time. 
The quasistar will then move rightwards in the  $M_*-M_{\rm BH}$ place and it will eventually cross the threshold growth line. 
Afterwards, the envelope mass will decrease abruptly until the quasistar will hit the no-solution line and the envelope will disperse.

The maximum mass for the BH is therefore given by eq.\ \ref{eq:tev=tacc}. 
This implies that it is {\em necessary} to have $M_* > 10^6 M_{\sun}$ in order 
to form a BH seed, more massive than $\sim 10^3 M_{\sun}$. 
We will discuss the implications of our findings in \S\ref{sec:discussion}, 
after having verified this picture with numerical calculations.

\section{Numerical Solution}
\label{sec:numerical}
 
 The full modeling of a quasistar, including the radiatively driven wind, cannot be carried out entirely analytically.
 In particular, one cannot derive the wind strength and its optical depth 
 as a function of $M_*$ and $M_{\rm BH}$. In part, this is because the stellar ``boundary" conditions 
 at the sonic point depend on the characteristics of the wind, and vice versa.
 
  In the previous section, this ignorance was buried
  in the unknown behaviour of the parameter $\Aconst$, and 
 of $\Gamma(r_{\rm tr})$ that enters the equations through $\kappa_{\rm tr}$.
 We simply assumed that their evolution was such that the threshold-growth line lies in the $\epsilon=1$ region.
 However, in order to determine the actual extent of the evaporation strip, it necessary to calculate numerically the structure of the quasistar.
 
 \subsection{The innermost accretion region}
 As was the case in the analytic analysis, we do not model the accretion flow.
 We assume that a luminosity $L_{\rm BH}$ given by eq.\ \ref{eq:Lbh_num}
 is transported through the quasistar. 
 
 \subsection{The Hydrostatic Convective Envelope}
\label{sec:hydrostatic_envelope}
Above the accretion zone resides the hydrostatic region of the envelope. It is composed of two parts. 
The inner part is the convective zone, in which the energy emitted by the accretion is convected outwards. 
Convection is an efficient process of energy transfer only as long as the luminosity $L_{\rm BH}$ is 
lower than the maximal convective luminosity, $L_{\rm c,max}$, given by 
eq.\ref{eq:lmax_c}.

The hydrostatic envelope is described by the equation of hydrostatic equilibrium,
\begin{equation}
\frac{1}{\rho}\frac{dP}{dr}=-\frac{GM(r)}{r^2},
\label{eq:hydro_eq}
\end{equation}
where $M(r)$ is the total mass contained within radius $r$,
\begin{equation}
M(r)=M_{\rm BH}+\int_{r_{\rm b}} ^r 4\pi r^2 \rho dr,
\label{eq:mtot}
\end{equation}
the equation of state,
\begin{equation}
P=P_{\rm g}+P_{\rm r}=\frac{k_{\rm b}T\rho}{\mu m_{\rm p}}+\frac{1}{3}aT^4,
\label{eq:P}
\end{equation}
where $k_{\rm b}$ is the Boltzmann constant and $\mu m_{\rm p}$ is the mean mass per particle.
 Given our primordial composition and that in most of the quasistar $T > 10^4$ K, we adopt $\mu =0.5$, independently of radius.
The last assumption does not affect our global results for two reasons. First, quasistars are radiation pressure dominated. Second, the pressure is important only over the hydrostatic region and the base of the wind, where the temperature is always above the ionization temperature.  
The temperature gradient is given by the adiabatic gradient,
\begin{equation}
\frac{dT}{dr}=\frac{\Gamma_2-1}{\Gamma_2}\frac{T}{P}\frac{dP}{dr},
\label{eq:dt_ad}
\end{equation}
where $\Gamma_2$ is the adiabatic index:
\begin{equation}
\Gamma_2=\frac{32-24\beta-3\beta^2}{24-18\beta-3\beta^2},
\label{eq:Gamma2}
\end{equation}
and $\beta = P_{\rm g}/P$.

\subsection{The Hydrostatic Radiative Region}
\label{sec:radiative layer}

 According to common wisdom, objects with a mass $M_*$ cannot shine beyond their classical
Eddington limit, $L_{\rm Edd} = 4 \pi GM_*c / \kappa$,  since no hydrostatic solution exists.  In
other words, if objects do pass $L_{\rm Edd}$, they are highly dynamic.
They have no steady state, and a huge mass loss should occur since their
atmospheres are then gravitationally unbound.  Thus, persistent astrophysical objects can
pass $L_{\rm Edd}$ but only for a short duration corresponding to the time
it takes them to dynamically stabilize, once super-Eddington conditions arise.

For example, this can be  seen in detailed 1D
numerical simulations of thermonuclear runaways in classical nova eruptions, which can achieve super-Eddington luminosities but only for several dynamical timescales \citep[e.g.,][]{Starrfield1989}. However, once they do stabilize, they are expected and indeed do reach
 in the simulations, a sub-Eddington state.  Namely, we
naively expect to find no steady state super-Eddington
atmospheres. This, however, is not the case in nature, where nova eruptions are clearly super-Eddington for durations which are orders of magnitude longer then their dynamical timescale \citep{ShavivNovae}. This is exemplified with another clear super-Eddington object---the great eruption of $\eta$-Carinae, which was a few times above its Eddington luminosity for over 20 years \citep{ShavivEta}.

The existence of a super-Eddington state can be naturally explained, once we consider the following: 
\begin{enumerate}
\item Atmospheres become unstable as they approach the Eddington limit. In addition to instabilities that operate under various special conditions (e.g., Photon bubbles in strong magnetic fields, \citealt{Bubbles1,Bubbles2,Blaes2001,Bubbles3}, or s-mode instability under special opacity laws, \citealt{Opacity1,Opacity2}), two instabilities operate in Thomson scattering atmospheres \citep{ShavivInstabilities}. It implies that {\em all atmospheres will become unstable already before reaching the Eddington limit}. 
\item The effective opacity for calculating the radiative force on an inhomogeneous atmosphere is not necessarily the microscopic opacity. Instead, it is given by
\begin{equation}
\label{eq:effectiveOpacity}
    \kappa_{V}^{\mr{eff}} \equiv {\left\langle F\kappa_{V}\right\rangle_{V}
      \over  \left\langle F \right\rangle_{V}},  
\end{equation}
where $\left< ~\right>_V$ denotes volume averaging and $F$ is the flux \citep{ShavivPorous}.
 The situation is very similar to the Rosseland vs.\ Force opacity means used in non-gray atmospheres, where the inhomogeneities are in frequency space as opposed to real space. For the special case of Thomson scattering relevant here, the effective opacity is always reduced.
\end{enumerate}


For $\Gamma \ge \Gamma_{\mr{crit}}$, the effective opacity can be described by an empirical parametrization \citep[e.g.,][]{Dotan2010}
\begin{equation}
\kappa_{\mr{eff}}=\kappa \left(1-\frac{A}{\Gamma^B}\right){ 1 \over \
\Gamma},
\label{eq:kappa_eff}
\end{equation}
where $\Gamma_{\mr{crit}}$ is the critical $\Gamma$ above which inhomogeneities are excited.
From theoretical considerations, we take $\Gamma_{\mr{crit}}= 0.8$ \citep{ShavivInstabilities}. In our work we choose $B=1$, which 
implies $A=0.16$.  

Finally, we assume that the relation between the effective Eddington factor $\Gamma_{\mr{eff}} \equiv L/L_ \mr{edd,r}(\kappa_{\rm eff})$ and 
the classical Eddington factor $\Gamma \equiv L/L_ \mr{edd,r}(\kappa)$ is empirically given by
\begin{eqnarray}
\Gamma_{\mr{eff}} & = & 1-\frac{A}{\Gamma^B}~~{\mr{for}}~~\Gamma >\Gamma_{\mr{crit}},  \nonumber \\
\Gamma_{\mr{eff}} & = & \Gamma~~{\mr{for}}~~\Gamma < \Gamma_\mr{crit}.
\end{eqnarray}
\no
As was found in \cite{Dotan2010}, the solutions we find in \S\ref{sec:results} are relatively insensitive to the exact choice of $A$ and $B$. 

The central temperature of a quasistar is of the order of $10^6$ K (see eq.\ 8) and it decreases slowly in the convective envelope. Therefore, the temperature in the hydrostatic part of the quasistar is high enough ($ \gg 10^4$K),  that the opacity is mainly due to electron scattering, and therefore approximately constant, $\kappa \approx \kappa_{\rm es}$.

Unlike the hydrostatic part, the temperature in the wind can drop below $10^4$ K, especially near the photosphere. This means that hydrogen recombines and opacities due to ``bound-free" transitions become important. Since we are modeling objects that are formed out of pristine or mildly polluted gas, we use the opacity computed by \citet{MD05}. In particular, we use an analytical fit

\be
\kappa = \frac{\kappa_{\rm es}}{1+\left({T}/{8 \times 10^3 {\rm K}} \right)^{-13}} \; {\rm cm^{2} \, g^{-1}}
\ee
\no
(BRA08), which captures the rapid drop in opacity for $T<10^4$K.

The temperature gradient is given by the radiative one with the above effective opacity,
\begin{equation}
\frac{dT}{dr}=-\frac{3\kappa_ \mr{eff}\rho L_{\rm BH}}{16\pi a c r^2 T^3}.
\label{eq:dt_rad}
\end{equation}
We note that in the atmosphere, the radiative luminosity is constant and $L=L_{\rm BH}$, because work has not been done on the gas.
In fact, the atmospheres of super-Eddington  objects, effectively remain sub-Eddington while being classically super-Eddington. 
This is true only as long as the inhomogeneties comprising them are optically thick. 
This condition will break at some point where the density is low enough and the wind will ensue.

\subsection{The Wind}
\label{sec:wind}
When inhomogeneities in the atmosphere becomes optically thin, the opacity returns to its microscopic value, and a net outward force acts on the gas.
We will show in the following, that the nature of the outflow and the mass loss rate ultimately depends on the ratio of gas to radiation pressure and therefore
on the quasistar mass $M_*$ (eq.~\ref{eq:beta}). 

\subsubsection{Two-fluid solution}
\label{sec:2-fluid_solution}
In a wind, the radiation energy can be transported by diffusion and by advection, so that the total energy rate is $L + \dot{M}_{\rm w} 4 P_{\rm r}/\rho$,
where the latter is the advected enthalpy, while $L$ is the diffusion luminosity,
\begin{equation}
L(r)=-\frac{dT}{dr} \frac{ 16\pi ac r^2 T^3}{ 3\kappa \rho}.
\label{eq:dTwind}
\end{equation}

In a low mass quasistar---we will quantify it shortly---the mean density is too low for advection 
to be efficient and the main mechanism for radiation energy transfer in the wind is 
diffusion.  The radiation can thus be described as an additional {\em external} force acting on the gas,
besides gravity. For this reason we refer to this solution as a ``two-fluid" wind.

Using the continuity and Euler equations in steady state, it is possible to show such trans-sonic wind solutions will be characterized by having an isothermal sonic point which coincides with the critical point. The ``isothermal" sonic point is the radius where $v = c_{\rm w} = \sqrt{P_{\rm g}/\rho}$ (with $c_{\rm w}$ being the isothermal speed of sound), while the critical point is the radius where the net forces balance each other. 

Using the above definition of the sonic point, the mass loss rate can we written as
\begin{equation}
\dot{M}_{\mr{w}}=4 \pi r_{\rm s}^2 \rho_{\mr{w}}c_{\rm w}=\mathit{const.},
\end{equation}
where, $\rho_{\mr{w}}$ is the density at the sonic point.

In our case, the critical point will be located where the net radiative force on a gas element balances gravity, namely, where $\Gamma_{\rm eff} \approx 1$. This will happen when the  inhomogeneities comprising the porous atmosphere become optically thin such that they cannot reduce the opacity anymore. This point will also mark the outer extent of the hydrostatic part of the quasistar, $r_{\rm s} \simeq r_*$.

Based on the fact that the inhomogeneities are the result of radiative hydrodynamic instabilities which operate on length scales comparable to the density scale height in the atmosphere, it is possible to estimate the average density at the sonic point \citep{ShavivNovae}. Using this density, the mass loss can be estimated to be 
\begin{equation}
\dot{M}_{\rm w,0} = {\cal W}\,L_{\rm BH} \frac{\left(1-\Gamma_{\rm s}^{-1}\right)}{ c_{\rm w}c},
\label{eq:mdotwind2}
\end{equation}
where 
$\Gamma_{\rm s}=\Gamma(r_{\rm s})$ and ${\cal W}$ is a dimensionless wind function \citep{ShavivNovae}. 
In principle, ${\cal W}$ can be calculated ab initio only after the nonlinear state
of the inhomogeneities is understood.  This however is still lacking
as it requires elaborate 3D numerical simulations of the nonlinear
steady state. Nevertheless, one can compare the above prediction to observed mass loss rates in classical novae and in giant eruptions of Luminous Blue Variables, and find that ${\cal W} \sim 5 - 15$ \citep{ShavivNovae}. We shall take a value of $10$.  By comparing eq.\ \ref{eq:mdotwind} and eq.\ \ref{eq:mdotwind2}, we have an expression for the efficiency factor $\epsilon = W v_{\rm esc}^2/(2 c c_{\rm w}) \left(1-\Gamma_{\rm s}^{-1}\right) \approx 5 v_{\rm esc}^2/ (c c_{\rm w})$.

As mentioned before, when the mass of the quasistar increases, the compactness of the star increases as well.
As a consequence, an increasingly high efficiency $\epsilon$ is necessary for the 
predicted mass loss to be pushed to $r\rightarrow \infty$ by the available luminosity. 
When the required efficiency satisfies $\epsilon >1$, which means $v_{\rm esc} \gtrsim \sqrt{c_{\rm w} c / {\cal W}}$,
a time independent wind solution is then not possible, because the wind would have to stagnate at a finite radius. 
The regime is then that of ``photon-tired winds'' \citep{Owocki}.

The time-dependent behavior of photon tired winds was numerically investigated by \cite{vanMarle}. 
It was found that shocks form between infalling material and the outflowing wind. 
This forms a  layer of shocks in which there is a large kinetic flux, but without the associated mass flux. 
In particular, they find that the mass loss from the top of the layer of shocks is reduced to $\epsilon \lesssim 1$:
$\epsilon \approx \min[0.2 \, \Gamma_{\rm s}^{0.6}, 0.9]$.

Therefore, in general, the wind mass loss is given by 

\begin{equation}
\frac{\dot{M}_\mr{w}}{\dot{M}_\mr{w,max}} \simeq \mr{min} \left(0.2 \, \Gamma_{\rm s}^{0.6}, 0.9, \frac{\dot{M}_{\rm w,0}}{\dot{M}_\mr{w,max}} \right).
\label{eq:mdot_wind_tot}
\end{equation}

Beyond the sonic point, acceleration to, say, $v \simeq 2 c_{\rm w}$ happens in a geometrically thin layer of a few scale heights.
In the following supersonic regime, the thermal gradient is no longer important 
with respect to the external forces. From this point on the gas velocity is described by

\begin{equation}
v\frac{dv}{dr}=-\frac{GM_{\rm T}\left[1-\Gamma(r)\right]}{r^2}.
\label{eq:vwind}
\end{equation}

The last equation that is needed to describe the wind is that of energy conservation, which for $r > r_{\rm s}$ reads

\begin{equation}
L_{\rm BH} =L(r)+\dot{M}_{\rm w}\left[\frac{1}{2}v^2-\frac{GM_{\rm T}}{r_{\rm s}}+\frac{GM_{\rm T}}{r} \right] = {\rm const},
\label{eq:Lwind}
\end{equation}
where the advection term has been consistently dropped.

We are now in the position to quantify the regime in which this ``two-fluid solution'' can apply. 
The requirement that the gas flows through a sonic point, beyond which $\Gamma > 1$ is equivalent to
the requirement that the advective term must not reduce the luminosity below the Eddington value by $r_{\rm s}$,

\begin{equation}
\dot{M}_\mathrm{w} {4 P_\mathrm{r} \over \rho} < L_\mathrm{\rm BH} - L_\mathrm{\rm Edd},
\label{eq:ad0}
\end{equation}   
This criterion will be slightly different depending on the mass loss regime, which determines $\dot{M}_{\rm w}$, but in both cases
it can be expressed in terms of a minimum gas to radiation pressure ratio.
In non-photon-tired winds, (eq.\ \ref{eq:mdotwind2}),  it becomes
\begin{equation}
\bar{\beta}  >  4 {\cal W}{c_{\rm w} \over c}.
\label{eq:non-PTAC}
\end{equation} 
We remind here that $c_\mathrm{w}$ is the isothermal sound speed calculated at the sonic point. 
In the photon-tired regime, the criterion for having an isothermal critical point is instead,
\begin{equation}
\bar{\beta} > {8 \Gamma \over \Gamma-1} \left( c_w \over v_{\rm esc} \right)^2.
\label{eq:PTAC}
\end{equation}

Because generally one has both that $c_w \ll c$ and $c_w \ll v_{\rm esc}$, the two-fluid solution with an ``isothermal" 
sonic point breaks down only for very low gas pressure fractions, which in quasistars implies 
a large envelope mass. Numerically, the line above which there is no isothermal critical point 
is denoted in fig.\ \ref{fig:menv_mbh_plot}. One can see that except for the least massive black holes, 
this line resides in the photon tired regime. This is because the non-photon-tired criterion (eq.\ \ref{eq:non-PTAC}) 
is generally more stringent than the photon tired criterion (eq.\ \ref{eq:PTAC}).
Eq.\ \ref{eq:non-PTAC} implies a mass range

\begin{equation}
\frac{M_*}{10^6 M_{\sun}} < 0.3 \left(c_{\rm w} \over 10^{7} {\rm cm/s}\right)^{-1/2}.
\end{equation}

\subsubsection{Adiabatic Winds}
\label{sec:adiabatic_winds}

At low enough gas pressures, no isothermal critical point can develop since the luminosity in the fluid
frame of reference is sub-Eddington.
Instead, the radiative energy is mostly advected with the flow. 
One can then consider the radiation and gas to flow together as one fluid, 
with an effective adiabatic index of the combined fluid (which is very close to 4/3). 

In this regime, the quasistar will develop a slow outflow above the porous 
atmosphere, enough to keep the flux sub-Eddington, but still sub-sonic. 
The flow would only become supersonic further out, where the escape velocity 
is similar to the local speed of sound, resulting in an adiabatic wind (i.e., the classical wind solution for stars, \citealt{parker:60}).

To see that such a subsonically outflowing layer must be present, let us assume for a moment that it is not. 
In such a case, an adiabatic ``single fluid" wind will develop with a sonic point located where the polytropic part of 
the quasistar ends (eq.\ \ref{eq:rqs}). To find the energy required to drive a wind from that location, we need to consider 
that the polytropic object has $P = K \rho^{4/3}$ with $K = (M/8\pi)^{2/3} \pi G$, that its radius is given by eq.\ \ref{eq:rqs}, 
and that for a thermal wind, $v \sim v_{\rm esc}$ at the sonic point. The ratio between this energy and the energy released by the black hole is then
\begin{equation}
{L_{\rm wind} \over L_{BH}} = {2 \pi r_*^2 \rho v_{\rm w}  v_{\rm esc}^2 \over L_{\rm BH} } \approx 12 \left(M_* \over M_{\rm BH} \right)^2 \gg 1.
\end{equation}
Clearly, the required energy is much larger than the amount available. This implies that the
thermal wind cannot be accelerated from where the polytropic part of the star ends. 
Therefore, there must be a layer with an outflow which is fast enough to keep the luminosity sub-Eddington 
(relative to the gas), but slow enough to be subsonic. Only further out, where the escape speed is significantly slower, will the flow become supersonic.  

Since the amount of energy left in the flow at the sonic point is much smaller than the 
energy it had deep below, it is clear that almost all the energy is used to pull the material 
out of the gravitational potential well, and very little will be left as kinetic energy in the thermal wind, or in the form of observable luminosity. 

Thus, because the system is limited by the gravitational well, adiabatic winds are characterized by the similar mass loss rate $\dot{M}_\mr{w,max}$ of photon tired winds, albeit in a configuration which is markably different from before. 

\subsection{Numerical Methods}
\label{sec:num_meth}
Practically, we perform the numerical calculation of a quasistar as follows.
\begin{enumerate}
\item Choose a black hole mass, $M_{\rm BH}$.
\item Choose an inner total pressure, $P_{\rm c}$.
\item {\em Guess} the gas to total pressure ratio at the inner radius, $\beta_{\rm c}=P_{\rm g}/P_{\rm c}$.
\end{enumerate}
With the above values, one can calculate the inner temperature, density, adiabatic speed of sound, Bondi radius and luminosity by respectively using
\begin{equation}
T_{\rm c}=\left[\frac{3(1-\beta_{\rm c})P_{\rm c}}{a}\right]^{1/4},
\end{equation}
\begin{equation}
\rho_{\rm c}=\beta_{\rm c} P_{\rm c}\frac{\mu m_H}{kT_{\rm c}},
\end{equation}
\begin{equation}
c_{\rm c}=\sqrt{4P_{\rm c}/3\rho_{\rm c}},
\end{equation}
\begin{equation}
r_{\rm b}=\frac{GM_{\rm BH}}{2c_{\rm c}^2}.
\end{equation}
The luminosity $L_{\rm BH}$ is given by eq. \ref{eq:Lbh_num} with $\alpha =1$.
In fact, $L_{\rm BH}$ is evaluated at  $5 \times r_{\rm b}$, 
to avoid the density cusp generated by the point mass potential of the BH (see eq. \ref{eq:hydro_eq}).

Using the above values, we integrate the equations of stellar structure for a convective envelope (eqs. \ref{eq:hydro_eq} - \ref{eq:dt_ad}) until
$L_{\rm BH} = L_{\rm c,max}$. When $L_{\rm BH} > L_{\rm c,max}$, we use the same equations except that the temperature gradient is instead given by eq. \ref{eq:dt_rad}. This integration is carried out up to $\kappa_{\rm eff}=1$.
This location, where $\Gamma_{\rm eff}\approx 1$, 
is the wind sonic radius $r_{\rm s}$ in the two-fluid regime (sec. \ref{sec:2-fluid_solution}). From this point outward, we integrate the wind equations (eqs.\ \ref{eq:vwind}, \ref{eq:mdot_wind_tot}
and \ref{eq:Lwind}), with the following initial conditions.
At $r=r_{\rm s}$ we take the value of the gas isothermal sound speed $c(r_{\rm s})$ and temperature $T(r_{\rm s})$ given 
by the hydrostatic solution and we set the initial wind sound speed, velocity and temperature to be $c_{\rm w} = c(r_{\rm s})$,
$v_{\rm w} = c_{\rm w}$ and  $T_{\rm w} = T(r_{\rm s})$. Assuming mass continuity at $r_{\rm s}$ with the wind rate given by
eq. \ref{eq:mdot_wind_tot}, we derive the initial wind density $\rho_{\rm w} =\dot{M}_{\rm w}/4\pi r_{\rm s}^2 c_{\rm w}$. 
The way we connect the hydrostatic solution to the wind uses the fact that the acceleration happens in a very narrow radial region, and we approximate
it to happen at one radius, $r=r_{\rm s}$.

Finally, the wind equations are integrated up to the photosphere $r = r_{\rm ph}$, where the optical depth is $\tau=2/3$.
At $r_{\rm ph}$, the temperature $T(r_{\rm ph})$ should be such that one recovers consistent surface 
conditions for the radiation field:

\begin{equation}
T(r_{\rm ph})= \left( \frac{L(r_{\rm ph})}{4\pi r_\mr{ph}^2 \sigma} \right)^{1/4}.
\label{eq:tph}
\end{equation}
To enforce it, the value of $\beta_{\rm c}$ (our only free parameter) is iterated and, for each successive guess,
 the quasistar structure is recalculated until eq.\ \ref{eq:tph} is satisfied.

\begin{figure}
\begin{center}
\includegraphics[width=0.48\textwidth]{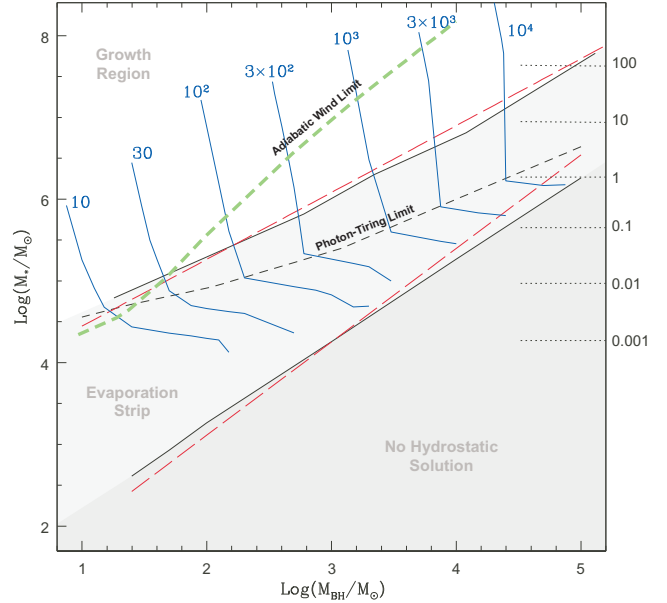}
\caption{The characteristic regions in the $M_*-M_\mr{BH}$ parameter space. Separated by black lines are: (1) The region with no hydrostatic solution (bottom) (2) The evaporation strip, where the envelope evaporation time-scale is shorter than the black hole growth time-scale (3) The growth region, where evaporation is less important. The short-dashed black line represents the threshold quasistar mass above which the winds are photon-tired. The blue lines represent lines of equal $\dot{M}_w$ [$M_{\odot}$ yr$^{-1}$].
The dotted lines are the horizontal equi-BH accretion rates, in units of $M_\odot$ yr$^{-1}$. The long-dashed red lines show the analytical predictions for the threshold-growth line and the no-solution line. The analytical threshold-growth line assumes a constant $\Gamma=10$, while the analytical no-solution line assumes constant $\Gamma=3$, $T_{\rm min}=4500$ K and the numerical $\kappa_{\rm eff}$.}
\label{fig:menv_mbh_plot}
\end{center}
\end{figure}

Since the adiabatic wind conditions are satisfied above the photon tiring limit (Fig. \ref{fig:menv_mbh_plot}), we do not need to calculate
specifically that solution, since the mass loss rate is going to be similar to the one of the corresponding photon tired solution. 


\section{Numerical Results}
\label{sec:results}

In this section, we first discuss the different regions in the $M_*-M_{\rm BH}$ parameter space for the structure of a quasistar. 
Then, we will follow its temporal evolution towards the formation of massive BH seeds.

\subsection{ The ``no-solution" region}
Using the numerical analysis, we first confirm the existence of a region in the $M_*-M_{\rm BH}$ parameter space where no hydrostatic self-gravitaing envelopes can be found. The region corresponds to $M_*/M_{\rm BH} \leq 18$, as depicted in Fig~\ref{fig:menv_mbh_plot}. 
In this region, the hydrostatic layer between $r_{\rm b}$, where the supersonic accretion begins, and $r_{\rm s}$, where a the supersonic wind starts, is simply too thin geometrically to be stable (with $r_{\rm s}/r_{\rm b}  \lesssim 2$).  

The boundary of this ``forbidden" region in the $M_*-M_{\rm BH}$ parameter space corresponds to envelopes with a photospheric temperature of $4500 \lesssim T_{\rm min} \lesssim 6000$ K, in agreement with BRA08. In particular, we find that $T_{\rm min}$ increases from $4500$ K to $6000$ K between $M_{\rm BH}=50 M_{\sun}$ and $M_{\rm BH}=10^4 M_{\sun}$; then it decreases down to $5400$ K at $10^{5} M_{\sun}$. For a given $M_{\rm BH}$, $T_{\rm min}$ is the minimum photopheric temperature with which a quasistar can shine, which corresponds to the minimum possible envelope mass, according to $T_{\rm ph} \propto M_*^{7/20}$ (eq.~\ref{eq:tph}).

We compare our numerical results with the analytic prediction given by eq.~\ref{eq:nsl} (lower long-dashed red line). It assumes a constant $\Gamma_{\rm s}=3$ and $T_{\rm min}= 4500$ K. The good agreement reinforces the validity of our numerical model. Despite the more sophisticated modelling of the atmosphere which allows for mass loss,  we recover the expected behaviour in the limit of low envelope masses. As the envelope mass decreases, for a given BH mass, $r_{\rm ph} \rightarrow r_{\rm s}$ and the wind becomes completely optically thin and dynamically unimportant.

\begin{figure}
\includegraphics[width=0.48\textwidth]{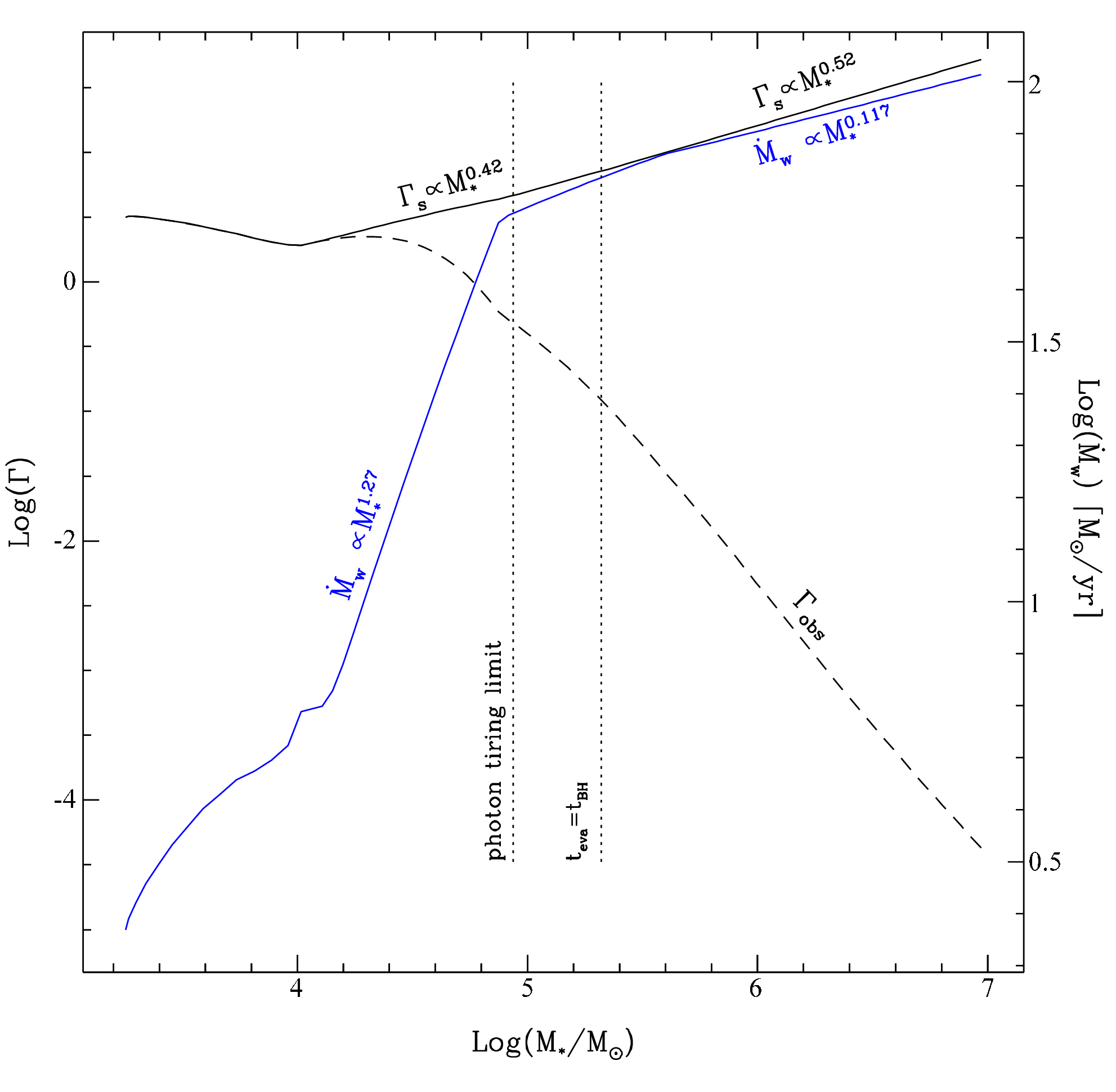}
\caption{Wind mass loss and Eddington ratios as a function of the envelope mass, for a fixed BH mass of $M_{\rm BH}=100 M_{\sun}$. The Eddington ratios are calculated at $r_{\rm s}$ ($\Gamma_{\rm s}$) and at the photosphere $r_{\rm ph}$ ($\Gamma_{\rm obs}\equiv \Gamma(r_{\rm ph})$).}
\label{fig:cons_mbh}
\end{figure}
   
\subsection{The Evaporation Strip}

At the no-solution line, quasistars are already very windy despite the relatively modest $\Gamma_{\rm s}$, with a loss of a few to ten thousand $M_{\sun}$ yr$^{-1}$, as shown by the lines of constant $\dot{M}_{\rm w}$ in Fig.~\ref{fig:menv_mbh_plot} (solid blue lines). There, the vigorous mass ejection causes the evaporation time-scale to be shorter
 than the accretion time-scale,  $t_{\rm ev}< t_{\rm BH} $.

The mass loss intensifies as a given BH increases its envelope mass, $\dot{M}_{\rm w} \propto M_*^{1.3}$ (Fig.~\ref{fig:cons_mbh}). It will eventually reach the photon tiring limit, after which it increases at a much slower rate  $\dot{M}_{\rm w} \propto M_*^{-1/5} \Gamma_{\rm s}^{3/5} \propto M_*^{0.1}$  (see eq.~\ref{eq:mdotwind} and Fig.~\ref{fig:cons_mbh}).  This can also be seen in Fig.~\ref{fig:menv_mbh_plot}, where above the short-dashed  line, the lines of constant $\dot{M}_{\rm w}$ become almost vertical. 
Correspondently, the evaporation time-scale first decreases as $t_{\rm ev} \propto M_*^{-0.3}$, then, once in the photon tiring regime, it {\em increases} as $t_{\rm ev} \propto M_*^{0.9}$.

The accretion rate onto the BH, instead, always increases linearly with the envelope mass. Therefore, the accretion time-scale goes as $t_{\rm BH} \propto M_*^{-1}$.
Because of this faster and monotonic evolution, it eventually equals the evaporation time-scale inside the photon tiring region.

The existence of an ``evaporation strip", where the envelope is lost to the wind rather than feeding the BH is thus confirmed by our numerical calculations (Fig.\ \ref{fig:menv_mbh_plot}). The extent of the strip, given by the line of equal time-scales ($t_{\rm ev} =t_{\rm BH}$) is well in agreement with our analytical prediction (eq.~\ref{eq:mdotwind}, upper red long-dashed line), where we use a constant $\Gamma_{\rm s} = 10$. As we explained in section \ref{sec:analytic_evolution}, we call this line ``the threshold growth line", since it indicates  the minimal envelope mass within which a given BH can significantly increase its mass by accretion before the envelope is lost.

\subsection{The evolution of a Quasistar}

So far, we focused on single equilibrium solutions. However, the structure depicted in Fig.~1 is in constant evolution. 
On the one hand,  the envelope 
loses mass to the wind and to the BH. On the other, it gains mass, accreting from the proto-galactic disc.
Since the dynamical time-scale $\sqrt{R_*^3/GM_*} \approx 20\ m_{\rm BH}^{6/5} m_{*}^{-4/5}  $ yr,  is always much shorter than both 
the accretion and the evaporation time-scales, we can construct the quasistar evolution as a succession of 
equilibrium states. 

 For this description to be valid, we require that the thermal timescale will be shorter than the evolutionary timescale. To estimate it, we should note that quasistars are to a good approximation $n=3$ polytropes, in which the internal energy almost compensates the gravitational binding energy, leaving a small contribution from the finite gas pressure. Namely, 
\begin{equation}
E_{\rm tot} \approx -\bar{\beta} E_{\rm grav}=\bar{\beta} k_2 G M^{5/3} \rho_c^{1/3},
\label{eq:Egas}
\end{equation}
where $k_2=0.639$. 
Since the thermal timescale is roughly $t_{\rm th} \approx E_{\rm tot}/L_{\rm Edd}$, we find using eqs.\ 4, 7 \& 10 that
\begin{equation}
t_{\rm th}<E_{\rm tot}/L_{\rm Edd}=2m_*^{7/10}m_{BH}^{-4/5} yr.
\label{eq:t_thermal}
\end{equation}  

As we shall see, this timescale is shorter or at most comparable to the evolutionary timescale. 

Once we are allowed to consider steady states, we can envision two scenarios. In the first scenario, we assume a large amount of mass collapses to form the quasistar, after which 
it equilibrates and evolves while the proto-galactic disk continues to be accreted. In the second scenario, 
we assume that the quasistar forms with a relatively low envelope mass, and it grows while the accretion 
rate is larger than the wind mass loss rate.

\begin{figure}
\includegraphics[width=0.48\textwidth]{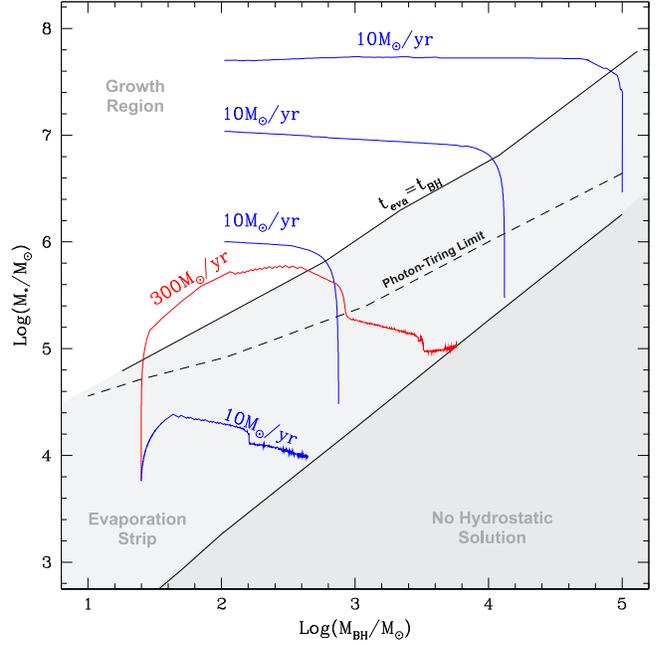}
\caption{Quasistar evolutionary tracks. The blue lines describe the evolution of quasistars which accrete from the pre-galactic disc at a rate of 
$10 \ M_{\sun}$ yr$^{-1}$. The upper three ones, have an initial BH mass of $100 M_{\sun}$, and initial envelope masses of $10^6, 10^7$ and  $5 \times 10^7 M_{\sun}$, respectively.
The quasistar with $10^6 M_{\sun}$ spends $1.5 \times 10^3$ yr in the accretion zone and a comparable time, $6 \times 10^3$ yr, in the evaporation strip. The aforementioned time-scales for
$M_* = 10^7 M_{\sun}$ (and $M_*= 6 \times 10^7 M_{\sun}$) are $1.3 \times 10^3$ yr (600 yr) and $1.5 \times 10^3$ yr (500 yr) respectively (see Fig.~\ref{fig:evolution}). In all cases, the mass evolution of the envelope is governed by the wind loses. In the BH growth zone, the quasistars lose only $\approx 25 \%$ of their initial mass, while all of it it is lost in the evaporation strip.
The final BH masses are: $750 M_{\sun}$, $1.3 \times 10^4 M_{\sun}$ and $10^{5} M_{\sun}$.
The lower blue line is for a quasistar fomed in the evaporation strip with $M_{\rm BH,i} = 25 \ M_{\sun}$ and $M_{\rm *,i} = 5.7 \times 10^3 M_{\sun}$,
while the red-line is for the same intial conditions but with $\dot{M}_{\rm acc} = 300 \ M_{\sun}$yr$^{-1}$.
In these last cases, the final BH mass is determined by accretion rates. There temporal evolution is shown in Fig.~\ref{fig:evolution}. }
\label{fig:menv_mbh_path}
\end{figure}

As study cases for the first scenario, we follow the fate of a BH of $100 M_{\sun}$ embedded within
envelopes of initial mass of $10^6$, $10^7$ and $7 \times 10^7~M_{\sun}$ (Fig.\ \ref{fig:menv_mbh_path}).
In each case, the quasistar is born in the accretion zone.
We considered several different accretion rates onto the envelope, $\dot{M}_{\rm acc} = 1,3,10,50  M_{\sun}$ yr$^{-1}$, and we found very similar tracks, since the change in $M_*$ is mainly due to wind losses, for the range of envelope and BH masses its track passes through. We therefore plot one examples with $\dot{M}_{\rm acc} = 10 \ M_{\sun}$ yr$^{-1}$. 

\begin{figure}
\includegraphics[width=0.48\textwidth]{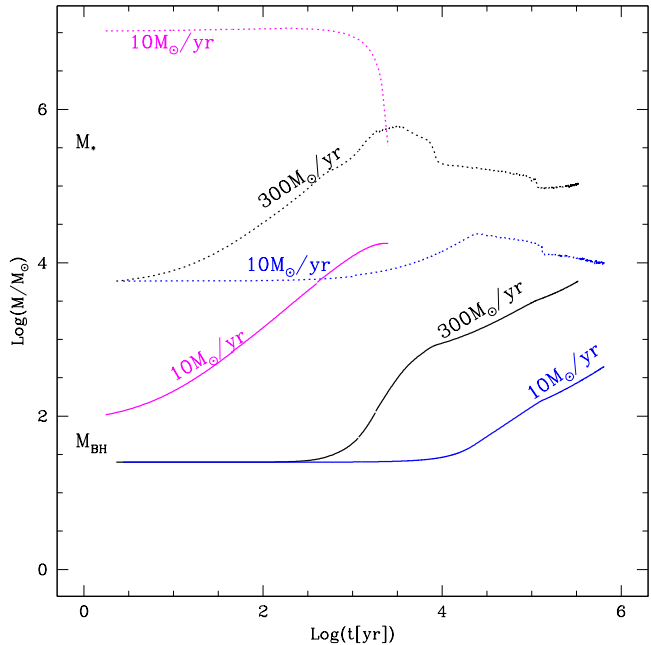}
\caption{The temporal evolution of the BH (solid lines) and envelope (dashed lines) masses for three quasistars. The magenta lines describe the case with $M_{\rm BH,i} = 100 M_{\sun}$ and $M_{\rm *,i} = 10^7 \ M_{\sun}$. Here the quasistar is formed in the growth region (see Fig.~\ref{fig:menv_mbh_path}) and reaches 
$M_{\rm BH,f} = 1.3 \times 10^4 M_{\sun}$, in $2.8 \times 10^3$ yr.  The blue and black lines describe the evolution of quasistars with an initial BH mass of $25 M_\odot$ and $M_{*,i} = 5.7 \times 10^3 M_\odot$, which is accreting at a constant rate of $10 M_\odot$ yr$^{-1}$ and $300 M_\odot$ yr$^{-1}$, respectively. The final BH formed in this scenario heavily depends on the accretion rate. It has a mass of $ 440 M_{\sun}$ in the first case and almost $5.7 \times 10^3\ M_\odot$ in the second case. Both quasistars live for $\approx 1$ Myr. Note that the thermal timescale in the latter case is comparable to the evolutionary timescale, and therefore it should be considered cautiously. }
\label{fig:evolution}
\end{figure}

As the BH mass increases by accretion, the quasistar moves towards the right of the plot, until it encounters the evaporation strip after a typical time $\approx  10^3$ yr has elapsed (see the temporal evolution in Fig.\ \ref{fig:evolution}).
This evolution in the accretion zone is so short, that the envelope loses only about $25 \%$ of its initial mass.
However, as soon as the quasistar enters the evaporation strip, the envelope is stripped off at such a high rate that the quasistar plunges down vertically in the diagram, reaching the no-solution line in a further $10^3 - 10^4$ yr.

The main result here is indeed that the presence of the evaporation strip limits the configurations in which an embryo BH can grow to masses $>10^3 M_{\sun}$. 
In particular, it is necessary that a quasistar  ``is born" above the line of threshold growth limit, with an envelope of at least $10^6 \ M_{\sun}$. The initial envelope mass determines the final BH mass. Empirically, for an initial BH mass of 
$100 \ M_{\sun}$ and $\dot{M}_{\rm acc} = 10 \ M_{\sun}$ yr$^{-1}$, we find

\begin{equation}
M_{\rm BH,f} \approx  760 \left( \frac{M_{*,\mr{i}}} {10^6 M_{\odot}} \right)^{1.25}  M_\odot,
\label{eq:finalBHmass_1}
\end{equation}

\no
where $M_{\rm BH,f}$ is the final BH mass and $M_{\rm *,i}$ is the initial enevelope mass. 
This relation holds for any reasonable accretion rate, $\dot{M}_{\rm acc}< 100 M_{\sun}$ yr$^{-1}$, 
and any initial BH mass $M_{\rm BH,i} \gtrsim 100 M_{\sun}$. For a much lower BH mass, the evaporation 
rate may be lower than the accretion rate and there may be a dependence on the initial BH mass,
likewise, for much higher accretion rates. 


In the second scenario, we assume that the quasistar is formed through a more gradual accretion process.
 For the ``low accretion rates", at least several dozen times the Eddington rate, a bare BH can handle the accreted mass 
with a disk like solution, having optically thick winds \citep{Dotan2010}. However, the BH will not be able 
to handle higher accretion rates with this configuration. This will necessarily puff up the system to form a ``minimal" 
quasistar. From this point, we wish to integrate its evolution. We expect the envelope mass to increase until the wind will 
compensate for the accreted matter. That is, it will evolve by increasing the mass of the cocooned BH, until the system will reach the no-solution line.

Two such tracks are plotted in Fig.\ \ref{fig:menv_mbh_path}, and the temporal evolution is given in Fig.\ \ref{fig:evolution}.
These quasistars are formed in the evaporation strip with $M_{\rm BH,i}= 25 M_{\sun}$. Here we find that the final mass of the 
BH is determined by the maximal envelope mass which in turn depends on the mass accretion rate. 

For $M_{\rm BH,i} = 25 \ M_{\sun}$, $M_{\rm *,i} = 1.4 \times 10^{4} M_{\sun}$, we calculate numerically relations between 
the accretion rates and the final BH mass. For accretion rates $< 20 M_{\sun}$ yr$^{-1}$, which do not allow the quasistar 
to leave the evaporation strip, we get the relation

\begin{equation}
{M_{\rm BH,f}} \approx 280 \left( \dot{M}_{\rm acc} \over 10 M_\odot/{\rm yr} \right)^{0.68}  M_\odot.
\label{eq:finalBHmass_2}
\end{equation}

For higher accretion rates which place the Quasistar above the growth line, we get

\begin{equation}
{M_{\rm BH,f}} \approx 550 \left( \dot{M}_{\rm acc} \over 10 M_\odot/{\rm yr} \right)^{0.62}  M_\odot.
\label{eq:finalBHmass_3}
\end{equation}

\no
Only in this last case, the final BH can be larger than $10^3 M_{\sun}$, but only for $\dot{M}_{\rm acc} \gtrsim 30 \ M_{\sun}$ yr$^{-1}$.
However, it is a weak function of $\dot{M}_{\rm acc}$, and already extremely high accretion rates ($> 100 \ M_{\sun}$ yr$^{-1}$) 
are needed to counterbalance the envelope evaporation and grow $M_{\rm BH} $ with at least a few $10^3 \ M_{\sun}$.  

In addition, we note that the case of an intial BH with a few tens of $M_{\sun}$ is a favourable configuration in which a quasistar 
formed in the evaporation strip can somewhat grow its BH. As we move towards the right within the evaporation strip, the winds become more and more vigorous (see Fig.\ 2) and the envelope is stripped off progressively more rapidly, leading eventually to conditions that will not allow any BH growth.

We thus conclude that, for plausible galactic accretion rates ($ \sim 1-10 \ M_{\sun}$ yr$^{-1}$), quasistars should be necessary formed {\em above} the
evaporation strip with an envelope of at least $10^6 M_{\sun}$, in order to get $M_{\rm BH} > 10^3 \, M_{\sun}$.
  
\subsection{Observational appearance}

During the growth of the BH, quasistars are very dim objects, since most of the radiation energy is converted into kinetic energy for the wind. The photons that do manage to escape from the photosphere, make up a modest luminosity which is smaller than the Eddington (see $\Gamma_{\rm obs}$ in Fig.~\ref{fig:cons_mbh}). 

When the quasistar enters the evaporation strip, however, the luminosity starts increasing fast as the envelope is being blown off. Eventually, $\Gamma_{\rm obs} \ge 1$ beyond the photon tiring limit, but this Super-Eddington phase is very brief ($<10^3$ yr), where most of the time is in fact spent at the photon tiring limit. 

Another interesting point is the fact that because of the wind, the objects never have a high effective temperature. As a consequence,  any cosmological redshift will imply that the objects can only be observed in the infrared band.  Therefore, we can conclude that although they were very brief transients,
they can in principle be detected by the forthcoming James Webb Space Telescope.

\section{Discussion and Conclusion}
\label{sec:discussion}

In the present work, we considered the state and evolution of quasistars, where a massive envelope enshrouds a black hole. We found that an important characteristic of these objects is the continuum driven winds which they accelerate. This gives rise to an evaporation strip in the envelope mass -- BH mass parameter space, where quasistars evaporate. BH growth takes place only for large enough $M_*$ for which the wind mass loss is limited by photon tired winds.

We then considered two evolutionary scenarios. In the first scenario, the quasistar is initially formed above the evaporation strip, by first having a large gas cloud collapse, then form an internal black hole, and subsequently evolve. In the second scenario, a seed BH is first formed, and then it accretes gas at large rates. 
We found that only the first scenario can generate large seed BHs. This is because once a BH is formed, unrealistically high 
mass accretion rates are required to evolve the quasistar out of the evaporation strip. 

These results put strong 
constraints on the dark matter halos in which massive SMBH seeds can form.
Our first scenario, on which we now concentrate, follows the picture put forth by \cite{Begelman2010}, whereby the progenitors of Quasistars are super massive stars that formed as the consequence of the high infall rate of hydrogen-cooled gas at the centre of dark matter halos.
The life-time of these stars is set by the thermonuclear timescale for burning their hydrogen core.  At the 
Eddington limit, it  is $\sim 2$ Myr, independent of mass. After that, the core may collapse into a BH. Once the BH starts accreting from the envelope, the feedback from the released luminosity  
sets up the structure we have investigated in this paper.

Our results constrain the initial mass of the quasistar, and thus of the supermassive star, that can grow massive SMBH seeds.
A quasistar with an initial mass $\approx 10^6 M_{\sun}$ (or $\approx 10^7 M_{\sun}$) can form a BH of $\approx 10^3 M_{\sun}$ (or $ \approx 10^4 M_{\sun}$ respectively, as can be see in Fig.\ 4).
This initial mass should be accumulated  in less than $\sim 2$ Myr, which requires accretion rates greater than a few 
$M_{\sun}$ yr$^{-1}$.

Dark matter halos accrete matter through their virial radius at a rate of   
\be
{\dot M}_{\rm DM} \approx  4.6  \ (z+1)^{2.5}_{10}  \left(\frac{M_{\rm h}}{10^9 M_{\sun}}\right)^{1.14}  {\rm M_{\sun} \, yr^{-1}},
\ee

\no
\citep{Neistein+06}, where $M_{\rm h}$ is the mass of the dark matter halo that hosts the quasistar and $(z+1)_{10} = (z+1)/10$.

Since the timescale for dark matter and gas accretion is the same \citep{Dekel+09}, 
\be
{\dot M}_{\rm acc} = f_{\rm b} \times {\dot M}_{\rm DM},
\label{eq:macc_cosmo}
\ee
it is also an estimate for the accretion rate that feeds the quasistar, under the assumption that almost all gas can 
be funnelled towards the halo centre. In eq.~\ref{eq:macc_cosmo}, $f_{\rm b} \approx 0.17$ is the cosmological baryon fraction \citep{komatsu+2010}.  

In fact, it is not clear if such high accretion rates do not undergo fragmentation at parsec scales  and if so, how much gas turns into stars (e.g., \citealt{Shlosman1989}; \citealt{Levin2007}). This is known as the ``active galactic nuclei (AGN) fueling problem", since we observe them shining with a luminosity that implies $\sim 1 M_{\sun}$ yr$^{-1}$. However if, like it must happen in AGNs, this problem is not severe \citep{Wise2008,Begelman2009}
 such flows can reach the proto-galactic centre.

Under this assumption, we can thus connect the accretion rate needed in order to form a quasistar of a certain mass and at a certain redshift, with the mass of the host halo through eq.~\ref{eq:macc_cosmo}.
However, we should also assume an efficiency factor, since not all gas $f_{\rm b} \times M_{\rm h}$ in the halo can be used to form a quasistar.
Therefore, the minimum halo mass that can host a given quasistar with mass $M_*$ is 

\be
\frac{M_{\rm h}}{M_{\sun}} \approx \max \left[\frac{7 \times 10^{8} }{(1+z)_{10}^{2.2}}\left(M_* \over10^6 M_{\sun}\right)^{0.9},6 \times 10^8 \frac{M_*}{10^6 M_{\sun}}\right],
\label{mh_min}
\ee

\no
where we assumed in the second term that no more than $1\%$ of the total amount of gas is used for the formation of the quasistar. The first term comes directly from eq.~\ref{eq:macc_cosmo}.

Eq.~\ref{mh_min} implies that at $z=10$, quasistars with $M_*>10^6 M_{\sun}$ can be found only in dark matter halos with $M_{\rm h} \gtrsim 10^9 M_{\sun}$. 
In particular, for $z \gtrsim 10$ ``massive" SMBH seeds of $M_{\rm BH} >10^4 M_{\sun}$ need host halos with $M_{\rm h} \gtrsim 6 \times 10^{9} M_{\sun}$.
 
 The very bright quasars observed at $z \sim 6$ have masses of typically $10^9 M_{\sun}$, estimated using the Eddington argument. Moreover, seeds of $10^4-10^5 M_{\sun}$ can grow 
 by accretion at the Eddington rate in $\sim 0.5$ Gyr. Therefore, such SMBH seeds may have formed as late as $z\sim 10$. At this redshift, the comoving number density of halos with $M_{\rm h} \approx 10^{12} M_{\sun}$  (calulated with the \citealt{ST99} formalism) matches the observed comoving number density of those bright quasars, $10^{-9} {\rm Mpc}^{-3}$ \citep{Fan2001}. Those halos are indeed above the threshold set by eq.\ \ref{mh_min}.

Of course, halos formed at different redshifts can lead to the observed high redshift quasars and to the observed local galaxies with SMBHs. To
properly derive the consequences of eq.\ \ref{mh_min}, one should thus follow a ``merging tree" evolution of halos and also of the 
hosted black holes. This is beyond the scope of this paper, but it will be addresses in a forthcoming work.

Finally, our results show that, unfortunately, Quasistars will not be easy to observe. 
They are relative rare  objects which shine at super-Eddington rates only for a short period, prior to their final evaporation.

\section*{Acknowledgements}

This research has been supported by the Israel Science Foundation, grant 1589/10.
We thank Avishai Dekel, Marcello Cacciato and Mitch Begelman for useful discussions.

\def\apjs{Ap.\ J.\ Supp.}

\bibliography{bibfile}{}
\bibliographystyle{mn2e}

\end{document}